\documentstyle[11pt,epsf]{article}

\def\void{}
\def\labelmark{}
\newenvironment{formula}[1]{\def\labelname{#1}
\ifx\void\labelname\def\junk{\begin{displaymath}}
\else\def\junk{\begin{equation}\label{\labelname}}\fi\junk}%
{\ifx\void\labelname\def\junk{\end{displaymath}}
\else\def\junk{\end{equation}}\fi\junk\labelmark\def\labelname{}}
{\ifx\void\labelname\def\junk{\end{array}\end{displaymath}}
\else\def\junk{\end{array}\right.\end{equation}}
\fi\junk\labelmark\def\labelname{}\def\junk{}
}
\newcommand{\beq}{\begin{formula}}
\newcommand{\eeq}{\end{formula}}
\newcommand{\beqv}{\begin{formula}{}}
\newcommand{\rf}[1]{(\ref{#1})}
\newcommand{\bea}{\begin{eqnarray}}
\newcommand{\eea}{\end{eqnarray}}

\begin{document}

\begin{titlepage}

\null
\begin{flushright}
 SU-4240-619
\end{flushright}
\vspace{20mm}

\begin{center}
\bf\Large A Real-Space Renormalization Group for Random Surfaces
\end{center}

\vspace{5mm}

\begin{center}
{\bf G. Thorleifsson}\\
{\bf S. Catterall}\\
Physics Department, Syracuse University,\\
Syracuse, NY 13244.
\end{center}

\begin{center}
\today
\end{center}

\vspace{10mm}
      
\begin{abstract}
We propose a new real-space renormalization group transformation for
dynamical triangulations. It is shown to preserve geometrical
exponents such as the string susceptibility and Hausdorff dimension.
We furthermore show evidence for a fixed point structure both
in pure gravity and gravity coupled to a critical Ising system.
In the latter case we are able to extract estimates for
the gravitationally dressed exponents which agree to within $2-3\%$ of
the KPZ formula.
\end{abstract}

\vfill

\end{titlepage}

\section{Introduction}

The success of dynamical triangulations (DT) in providing a discrete
regularisation for two dimensional quantum gravity is well known
\cite{gen}. The advantages of this approach are twofold; firstly
it allows powerful analytical techniques based on matrix models to be
employed and secondly it lends itself to non-perturbative studies
via numerical simulation.
The latter approach can be applied both to study models
which are currently not amenable to exact treatments (for example
with matter central charge $c>1$) 
and to compute quantities of physical interest, such
as the nature of the quantum geometry, which
are difficult or impossible to arrive at with analytical methods. 

Much of the numerical work has centered around the 
application of finite size scaling to extract critical 
exponents (see e.g.\ \cite{finite}).
In statistical mechanics the validity of such techniques 
depends on the existence of
a renormalization group (RG) which governs the flow
of couplings under changes in scale. It is tempting to
believe that a similar RG structure must underlie the DT models
and is responsible for the observed scaling.
In the continuum formulations of quantum gravity the very issue
of a renormalization group is a difficult one to formulate since the theory
(in the absence of a cosmological constant) possesses no length
scale. In contrast the DT formulation contains an invariant cut-off
corresponding to the elementary triangle edge length. The latter
may be traded in for the number of triangles $N$ if the physical
volume is held fixed.

A successful RG transformation would be important both conceptually
and as a powerful new tool with which to compute critical
points and critical exponents for systems 
coupled to quantum gravity. 
Recently various real-space RG transformations
for dynamical triangulations have been proposed.
The first of these
approaches, due to Renken \cite{renken}, achieves this 
by randomly selecting a subset of nodes on
a given triangulation.  These nodes are then 
connected by links to form coarser triangulations, in
such way as to preserve {\it locally} the relative
geodesic distances of the selected nodes.  We can call this a local
geodesic blocking (LGB).

Another approach advocated by Krzywicki et al.\ \cite{krz} 
constructs a coarser triangulation by eliminating a class 
of extremal baby
universes associated with the original random lattice
(extremal implies that no further baby universes grow
on them). Since the
baby universe distribution is intimately connected to the
fractal structure of the ensemble of triangulated manifolds,
it is hoped that this transformation will (approximately) 
preserve the fractal
structure. This may be termed a fractal blocking (FB).

In this paper we propose a new real-space RG transformation
for dynamical triangulations based on a {\it local} 
node decimation scheme (ND).  The essence of this method is that nodes
are removed randomly from a triangulation.  After each one
is removed its neighbors are reconnected in such a way as to 
preserve the integrated ''local`` curvature and maintain
the triangulated structure of the manifold.
As we will demonstrate this method preserves the relevant
long distance physics of the surfaces, and in addition allows
us to compute the critical behavior of matter fields living
on the ensemble of random manifolds.

The paper is organized as follows:  In section 2 we introduce 
the methodology of Monte Carlo renormalization group.
In section 3 we discuss the problems involved in
blocking dynamical triangulations and describe the
method we propose.  In section 4 we apply the method
to pure gravity, demonstrate that it preserves the fractal
structure and show evidence for 
a fixed point behavior. 
We also look at the effects of including irrelevant
operators in the action.
In section 5 we apply the method 
to a critical Ising model coupled to gravity and show that,
combined with an appropriate blocking of the Ising spins,
it yields the correct critical behavior.
In section 6 we discuss some alternative blocking schemes.  
Finally section 7 summarizes our results.

\section{Monte Carlo Renormalization Group}

In conventional lattice field theory a 
RG or block transformation acts so
as to reduce the number of degrees of freedom by replacing the
original (bare) theory by one defined on a coarser lattice. 
This effectively integrates out the short distance
fluctuations, allowing us to focus directly on the 
long distance physics which governs the critical
behavior.  Associated with this blocking there will be a flow
in the coupling constants of the system to compensate
for the change in length scale.  If the system is
close to a continuous phase transition this flow will pass
close to a critical fixed point. Studying the 
flow in the vicinity of such a critical point allows us to
determine the critical exponents of the model. 

Consider a system of spins $\{\sigma\}$ on the sites of
some regular lattice with interactions described by
a Hamiltonian $H(\sigma)$.  All thermodynamic quantities
can be found from a detailed knowledge of the partition function
\beq{*21}
Z(H) = \sum_{\{\sigma\}} \: {\rm e}^{\textstyle H(\sigma)}.
\eeq
We define a RG transformation that maps the system onto
a blocked system, with fewer spins $\mu$, by
\beq{*22}
{\rm e}^{\textstyle H^{\prime}(\mu)} = \sum_{\{\sigma\}} \;
P(\sigma,\mu) \; {\rm e}^{\textstyle H(\sigma)}.
\eeq
$P(\sigma,\mu)$ is a {\it projection} operator that
couples together the spins of the original and the blocked 
system.  Note that as the background lattice is regular,
the coarser lattice trivially has the same structure. 
 
If the RG transformation is to preserve the physics of
the model the projection operator
should satisfies the following conditions;
\beq{*23}
P(\sigma,\mu) \geq 0 \;\;\;\;\;\;\; \forall \;\; \sigma,\mu
\;\;\;\;\;\; {\rm and} \;\;\;\;\;\;\
\sum_{\{\mu\}} P(\sigma,\mu) = 1.
\eeq
The former condition ensures that the renormalized
Hamiltonian is  real,
the latter that
the partition function is preserved.
In addition it is usually required that
$P(\sigma,\mu)$ preserves the symmetries of the Hamiltonian.

Expanding the Hamiltonian on some
operator basis $H(\sigma) = \sum_{\alpha}
K_{\alpha} O_{\alpha}(\sigma)$ 
the definition of a RG transformation Eq.\ \ref{*22}
implies a set of recursion relations for the
couplings 
\beq{*24}
\vec{K}^{(k+1)} = R[\vec{K}^{(k)}].
\eeq
The critical behavior of the model is calculated by looking 
at fixed points of these recursion relations,
$\vec{K}^*=R[\vec{K}^*]$. Close to such a fixed
point the flow of the difference in the couplings from their
fixed point values may by analyzed in a linear
approximation; 
\beq{*26}
\delta K^{(k+1)}_{\alpha} \;=\; \sum_{\beta} \left.
\frac{\partial K^{(k+1)}_{\alpha}}{\partial K^{(k)}_{\beta}}
\right |_{\vec{K}=\vec{K}^*} \!\!\!\!\delta K^{(k)}_{\beta}
\;\equiv\; \sum_{\beta} T^*_{\alpha \beta} \delta K^{(k)}_{\beta} .
\eeq
The critical exponents are then determined from
the eigenvalue equation 
\beq{*27}
\sum_{\beta} T_{\alpha \beta} \; u^i_{\beta} 
\;=\; \lambda_i \; u^i_{\alpha}.
\eeq
We distinguish between three types of eigenvalues:
$\lambda_i > 1$ is called
a {\it relevant} eigenvalue, $\lambda_i = 1$ {\it marginal}
and $\lambda_i < 1$ {\it irrelevant}.  The meaning
is simple.  Consider expanding the
operators in the effective action on a basis
given by the eigenvectors of Eq.\ (6).  Then, under iteration
of the RG step, components along the direction of
an irrelevant eigenvector ($\lambda_i<1$) will
be driven to zero. Conversely, components along the
relevant directions will increase. These directions
then describe how the renormalized Hamiltonian flows away
from the fixed point.
The case of a
marginal eigenvalue is special and has to be studied closely
in any individual case. 

A relevant eigenvalue defines a scaling exponent $y_i =
\log{\lambda_i}/\log{b_l}$ which describes the singular
behavior of the free energy ($b_l = N^{(k)}/N^{(k+1)}$
is the volume rescaling).  It can be related to
a combination of critical exponents of the model.

Although the condition imposed on the projection 
operator Eq.\ \rf{*23} allows plenty of freedom
in defining a RG transformation, in practice  it is
very hard to find one that allows for an explicit
calculation of the linearized matrix $T_{\alpha \beta}$.
One solution is to combine an exact RG
transformation with Monte Carlo simulations 
- the Monte Carlo renormalization group.
Monte Carlo simulations
are used to generate a sequence of configurations
characteristic of the original bare Hamiltonian, which
then are blocked using the RG transformation.
If the transformation is `apt' this will result in a
configuration describing the same long distance physics,
but on a coarser lattice.
There are two main sources of error in such a calculation;
the statistical error arising from the finite number of
configurations generated by the Monte Carlo algorithm and the
necessary truncation of the operator basis for $T_{\alpha\beta}$.
The former can be estimated and systematically reduced, the latter
can be assessed by varying the operator basis.

The renormalized Hamiltonian is approximated
using a set of operators $O_\gamma$ 
on the finite lattice and
the linearized matrix $T_{\alpha \beta}$ calculated 
by solving a set of chain rule equations:
\beq{*28}
\frac{\partial <O^{(k+1)}_{\gamma}>}{\partial K^{(k)}_{\beta}}
\: = \: \sum_{\alpha}
\frac{\partial K^{(k+1)}_{\alpha}}{\partial K^{(k)}_{\beta}}
\frac{\partial <O^{(k+1)}_{\gamma}>}{\partial K^{(k+1)}_{\alpha}}
\eeq
where
\bea
\frac{\partial <O^{(k+1)}_{\gamma}>}{\partial K^{(k)}_{\beta}}
& = & <O^{(k+1)}_{\gamma}\;O^{(k)}_{\beta}>
- <O^{(k+1)}_{\gamma}><O^{(k)}_{\beta}>,   \\
\frac{\partial <O^{(k)}_{\gamma}>}{\partial K^{(k)}_{\alpha}}
& = & <O^{(k)}_{\gamma}\;O^{(k)}_{\alpha}>
- <O^{(k)}_{\gamma}><O^{(k)}_{\alpha}>.
\eea
The critical exponents are then calculated from the eigenvalues
of $T_{\alpha \beta}$.  One appealing feature of this
method is that it is relatively easy to include a large 
set of operators, allowing for a systematic improvement
of the approximation.

\section{Random surfaces}

The model we are interested in is the dynamical triangulation
formulation of $2d$ quantum gravity.  It is defined
as a sum over all possible ways of gluing equilateral
triangles together in such way that they form a
closed (piecewise linear) manifold $T$;
\beq{*31}
Z(\mu) \; = \; \sum_{T \in {\cal T}} \;
{\rm e}^{-\mu N} \; Z_M\left(T\right).
\eeq 
$\mu$ is the cosmological constant, $N$ the volume
of the triangulation and ${\cal T}$ a
suitable class of triangulations.
In this paper we chose this class to be
combinatorial two-manifolds, i.e.\ no two nodes
are connected more than twice and no node 
is connected to itself. This implies that
there are no nodes of order two or less. Finally
$Z_M\left(T\right)$ denotes the partition function corresponding to
a sum over all possible  
matter fields living on the triangulation $T$.

In numerical simulation it is more convenient to study
this model at fixed volume, i.e.\ to look at the 
micro-canonical partition function $Z(N)$  
related to the grand canonical partition function
Eq.\ \rf{*31} through a discrete Laplace transformation
\beq{*32}
Z(\mu) \; = \; \sum_N {\rm e}^{-\mu N} \; Z(N).
\eeq

To construct a blocking transformation for this model
is quite different than for usual field theory models on 
regular (flat space) lattices.  There the RG transformations
trivially preserve the lattice structure and the main
problem is to devise a blocking scheme for the matter fields.
In contrast
it is not possible to preserve exactly the features of
any given fine triangulation under blocking.  Indeed the
lattice itself now carries dynamical degrees of freedom.
Thus any transformation will have to replace a given
triangulation $T(N)$ by some triangulation $T^\prime(N^\prime)$,
corresponding to a blocking factor $b_l = N/N^\prime$,
\beq{*33}
T^\prime(N^\prime) \;=\; R[T(N)].
\eeq
Clearly the choice of the transformation $R$ is of crucial
importance.  Presumably it should have the property of 
preserving certain aspects of the long distance
geometry.  But what aspects?  We propose a very
simple method based on a local node decimation of the 
triangulation.
  
\begin{figure}
\centering 
\epsfxsize=4.7in  \epsfbox{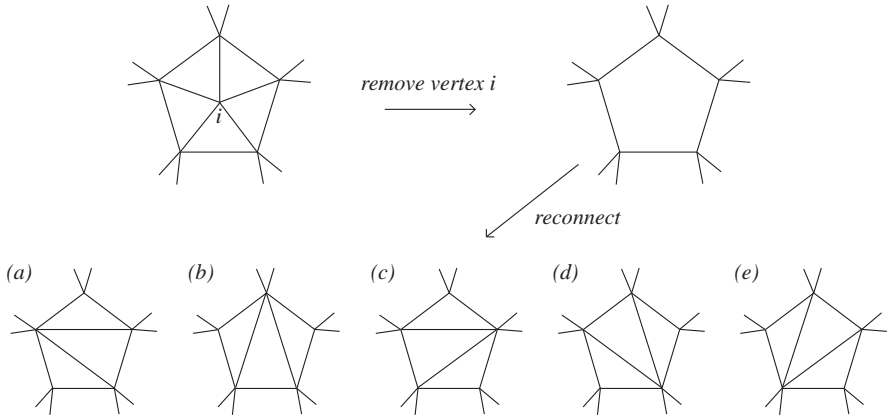} 
\caption{ The five possible ways of re-triangulate after
 removing a node of order five. 
 Choosing one of these possibilities
 at random yields a blocked triangulation.
 If a node of order six is removed there are 42
 ways of reconnecting. }
\label{fig:1}
\end{figure}

First pick a node at random in the
initial triangulation $T$.  If we attempt to
remove this node then unless its coordination is three
we will be left with a polygonal "hole" in the
triangulation.  However by suitable link addition 
it is possible to re-triangulate the interior of
this hole.  This can be done in many ways; we choose
one of those at random.  This is illustrated in 
Fig.\ \ref{fig:1}.

By iterating this procedure an arbitrary number of times
we can produce a blocked triangulation of any volume.  In
practice, this procedure is effected by randomly
flipping the links around the selected node until its
coordination number becomes equal to three.  At this
point any curvature associated with the node has been
smeared out over its neighbors.  We then remove the
three-fold node.  It is trivial to see that
this method preserves the {\it integrated} curvature within a
sufficiently large loop around the node in question. 
However it is not at all
obvious that this is enough to capture the most important
long distance physics governing the fractal structure,
critical exponents etc. We examine these issues now. 

\section{Pure gravity}

We have applied this blocking method to the case of
pure gravity (triangulations taken with weight unity)
and performed a series of
Monte Carlo simulations on lattices 
ranging from
250 to 4000 nodes.  A standard link flip
algorithm was used to explore the space of
configurations and typically $10^6$ sweeps 
performed for each lattice volume.  Every twenty 
sweeps a configuration was picked out and blocked
several times down to 8 nodes. 

\subsection{The fractal structure}

To see whether the long distance physics of the
surfaces is being preserved we have measured
two exponents which describe complementary features
of the fractal structure; the {\it string susceptibility
exponent} $\gamma_s$ and the {\it Hausdorff} or {\it fractal
dimension} $d_H$.

The string susceptibility exponent is related to
the large volume behavior of the fixed area
partition function $Z(N)$,
\begin{equation}
Z(N) \;\sim\; N^{\gamma_s - 3} \; e^{\mu_c N}.
\end{equation}
In the case of pure gravity $\gamma_s = -1/2$.
An efficient method for measuring $\gamma_s$ in numerical
simulations is to look at the distribution of so-called 
{\it minbu's}
(minimal neck baby universes) \cite{baby1,baby2}.
A minbu is a part of a triangulation which is
connected to the rest through a neck
or a loop consisting of three links\footnote{As we do not 
allow degenerate triangulations this is a minimal loop
on the surface.}.  The distribution of minbu's can be
calculated from the number of ways a baby universe of
volume $B$ can be glued onto a surface of volume $N-B$;
\beq{*42}
n_N(B) \;=\;  \frac{B\;Z(B)\:(N-B)\:Z(N-B)}{Z(N)}
\;\sim\; (N-B)^{\gamma_s - 2} \; B^{\gamma_s - 2}.
\eeq 
It is easy to measure this distribution numerically and
the exponent $\gamma_s$ is obtained by fitting to
Eq.\ \rf{*42}.  The accuracy of the fit can be improved
considerably by including a $1/B$ correction term.

\begin{figure}
\centering
\epsfxsize=4.7in \epsfbox{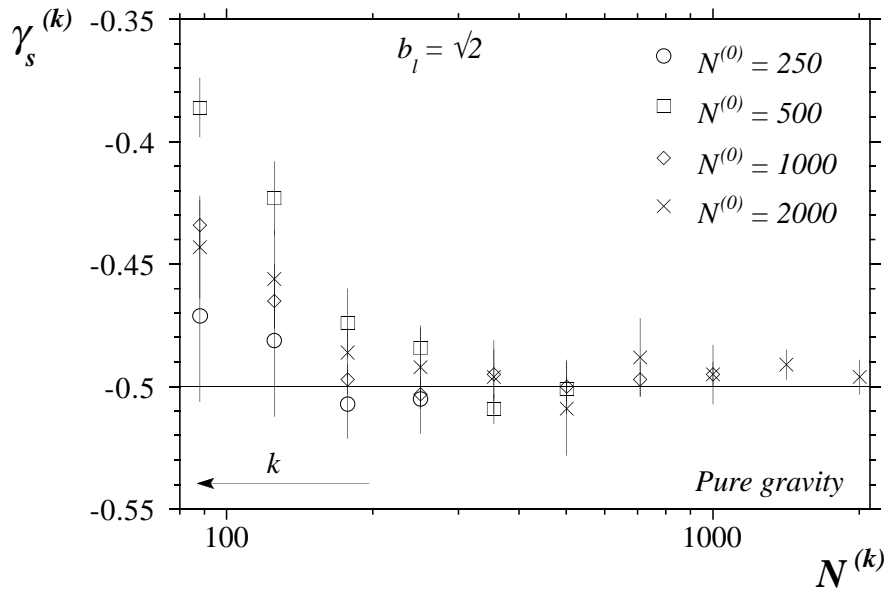}
\caption{$\gamma_s$ for ensembles
 of surfaces obtained by blocking pure gravity
 triangulations $k$ times.  The values are plotted versus
 volume and the blocking increases in the left direction. }
\label{fig:2}
\end{figure}

We have measured $\gamma_s$ in this way for the ensemble
of triangulations $T^{(k)}$ obtained by blocking $k$ times. 
The result is shown in Fig.\ \ref{fig:2} where we plot
$\gamma_s^{(k)}$ versus volume.  This is done for four different
initial volumes $N^{(0)}$.  It should be noted
that the number of blocking steps increases to the left
in the plot. 
It is clear from Fig.\ \ref{fig:2} that independent of the
initial volume and number of blocking steps
we get the correct value of $\gamma_s$ down
to volumes of size 200 nodes.  For smaller volumes
it is hard to measure $\gamma_s$ accurately
due to finite size effects.  
This demonstrates that 
even after removing 90\% of the initial  
triangulation by blocking 
the model remains in the universality
class of pure gravity. 

Additional evidence is obtained by measuring the
Hausdorff dimension of the surfaces.  It has recently been
shown that the Hausdorff dimension is
related to point-point distributions on the
triangulations and that it can be measured numerically
by looking at their scaling behavior
\cite{us,ajw}.  The point-point 
distributions $n(r,N)$ are defined as the number
of nodes at a geodesic distance $r$ from some
random origin.
Given that the system is close to criticality, it
is plausible and indeed appears correct, that these distributions satisfy 
a scaling ansatz for large $N$. This implies that
\beq{*45}
n(r,N) \;=\; N^y \; f\left(r/N^x\right) 
\:=\; N^{1-1/d_H} \: f\left({r\over N^{1/d_H}}\right). 
\eeq
The scaling exponents $x$ and $y$ are fixed
by assuming that there exist a 
power-law relation between a length scale $l$ 
and volume N,
$N \propto l^{d_H}$, and by noting that the integral
of $n(r,N)$ yields the total volume.

In numerical simulations $d_H$ can be extracted by
measuring $n(r,N)$ at different volumes and
find the optimal value that collapses
all the data onto a single curve using Eq.\ \rf{*45}.
The scaling assumption also
implies that both the position $r_0$ and height 
$n(r_0)$ of the peak in $n\left(r,N\right)$ 
should scale simply with volume $N$, i.e.
\beq{*46}
r_0 \sim N^{1\over d_H}  \;\;\;\;\;{\rm and}\;\;\;\;\;
n(r_0) \sim N^{1-{1\over d_H}}.
\eeq

In Table \ref{tab:1} we show $d_H$ measured using the 
relations Eq.\ \rf{*46} on volumes ranging from
250 to 1000 nodes.  This is shown both for the initial
(unblocked) lattices, and for ensembles obtained by
blocking once or twice.
It is clear that the
extracted Hausdorff dimension for the blocked ensembles is
completely in agreement with its value extracted from the bare
lattice $d_H=3.15(2)$.
But note that these results are far from 
the exact result $d_H=4$.
This effect has
been observed before and is attributed to the presence of
large corrections to scaling in the current lattice ensemble in
which degenerate links are excluded. In \cite{us} it was shown that
this estimate moves up to be statistically consistent with
four if a wider ensemble is used in which tadpole and
self-energy insertions are included.

\begin{figure}
\centering
\epsfxsize=4.7in \epsfbox{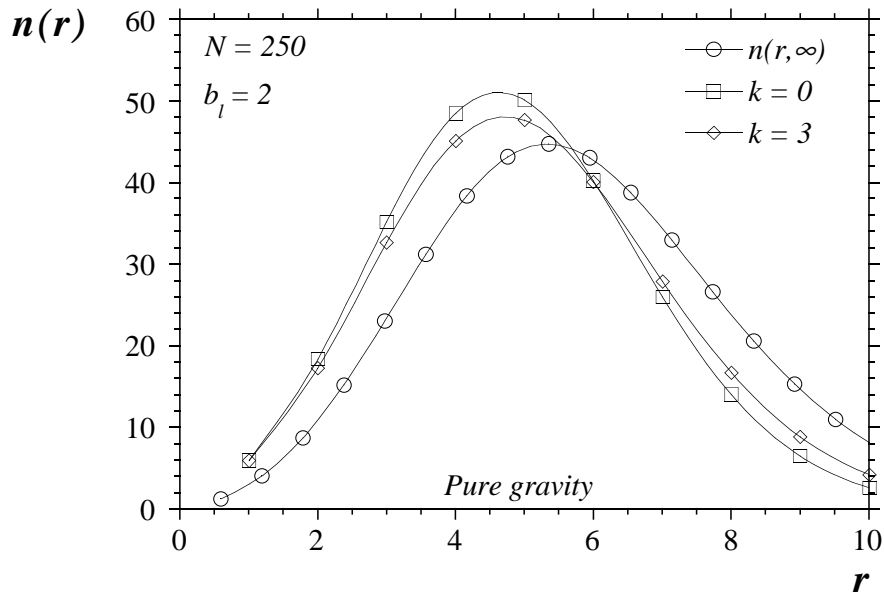}
\caption{The point-point distributions $n(r,N)$ for 
 $N = 250$ in the case of pure gravity.  Data for
 both blocked (87.5\%) and unblocked triangulations 
 is compared to an estimate of
 the continuum distribution $n(r,\infty)$.  }
\label{fig:3}
\end{figure}

It is more instructive to compare the distributions
directly.  In Fig.\ \ref{fig:3} we plot distributions
at volume $N = 250$, both for blocked and unblocked
triangulations. This is compared 
to an estimate of the 
infinite volume distribution $n(r,\infty)$. 
(the distribution $n(r,\infty)$ is obtained by scaling a 
distribution corresponding to a large volume ($N = 2000$)  
down to $N = 250$ using 
Eq.\ \rf{*45} with the exact value $d_H = 4$).
The finite size effects are very apparent in Fig.\ \ref{fig:3}
as both distributions deviate from  $n(r,\infty)$.
But it is intriguing that the
curve corresponding to the blocked $N = 250$ ensemble
is closer to the estimated 
continuum curve - perhaps the first indication that
the blocking is resulting in a flow towards a fixed point.

\begin{table}
\centering
\begin{tabular}{|cclcl|} \hline
$k$ && $d_H(r_0)$  &&  $d_H(n_{max})$  \\  \hline
0  &&  3.15(2)  &&  3.13(1)  \\
1  &&  3.16(2)  &&  3.15(1)  \\
2  &&  3.15(2)  &&  3.20(1)  \\ \hline
\end{tabular}
\caption{The Hausdorff dimension $d_H$ extracted both from the
 scaling of the location of the peak in $n(r,N)$
 and its maximal value.  Results are shown for different
 number of blocking steps $k$ with $b_l = 2$.
 The volumes used are from 250 to 1000 nodes.}
\label{tab:1}
\end{table}

Combined the measurements of $\gamma_s$ and $d_H$
for the ensemble of blocked manifolds provide strong
evidence that the fractal structure
is preserved.  This gives us confidence that the
DB method correctly captures the long distance 
physics of the model.

\subsection{Fixed point behavior}

It is not enough that the RG transformation preserves
the fractal structure, it should also induce a flow in
the couplings towards a critical (unstable) fixed point.
Optimally we would like to study the behavior of
relevant operators around that fixed point and
extract some critical exponents.
Unfortunately for pure gravity the only simple relevant
operator corresponds to the area operator, which is
not accessible to us as we are working in the fixed
area (volume) ensemble.  

Instead we have looked at the behavior of expectation 
values of various {\it irrelevant} operators constructed out of the local
coordination number $q_i$ or curvature $\left(6-q_i\right)$.
Specifically we have looked at the following operators (where
the superscript $k$ labels the blocking level):
\begin{eqnarray}
\left<O_1^{\left(k\right)}\right>& =& \left<{1\over N^{\left(k\right)}}
\sum_i\left(q_i^{\left(k\right)}
-6\right)^2\right>,\\ 
\left<O_2^{\left(k\right)}\right>& =& \left<{1\over N^{\left(k\right)}}
\sum_{\left<ij\right>}\left(q_i^{\left(k\right)}-6
\right)\left(q_j^{\left(k\right)}-6\right)\right>.
\end{eqnarray}
$\left<ij\right>$ implies that the sum is over neighboring nodes.

\begin{figure}
\centering
\epsfxsize=4.7in \epsfbox{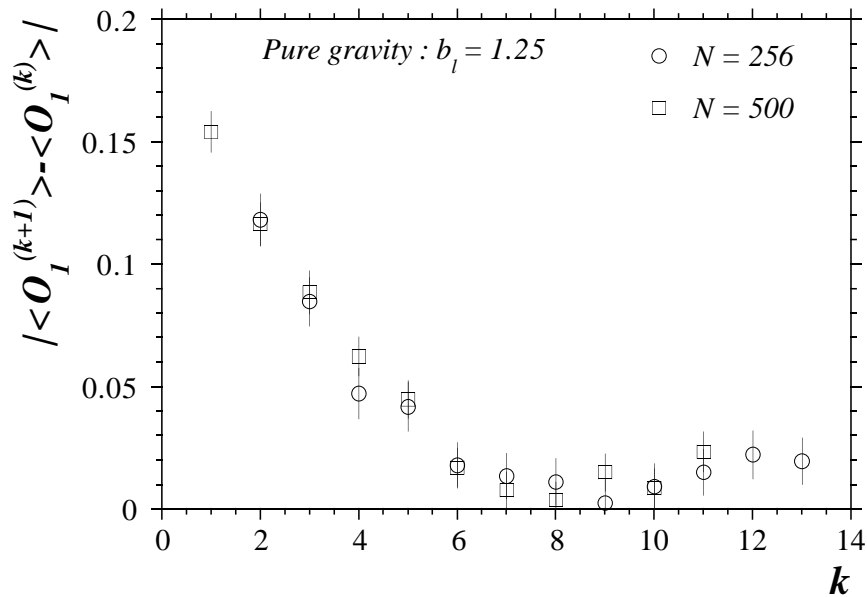}
\caption
 { Differences in expectation values of the
   operator $O_1^{(k)}$ for surfaces obtained
   with different amount of blocking (but with
   the same volume). }
\label{fig:4}
\end{figure}

If a fixed point exists then we
would expect that in the absence of finite size effects (i.e. the 
initial lattice is lying on the critical surface)
that these operators flow 
under repeated blocking into the fixed point. Thus differences
of the expectation values between blocking levels
should vanish;
\beq{*322}
\left|\left<O_i^{\left(k+1\right)}\right>
-\left<O_i^{\left(k\right)}\right>
\right|\to 0 \; ,\qquad k\to\infty.
\eeq
Since we are forced to work with finite volumes
we have to be careful in how we compare these operator
expectation values.
It only makes sense to compare operator expectation
values at the {\it same} volume, but differing in the number of
blocking steps. This we do in Fig.\ \ref{fig:4}.
We have simulated at different initial volumes $N^{(0)}$ 
(500 to 8000 nodes) which
then are blocked down to fixed volume (256 or 500 nodes) 
using a blocking factor $b_l = 5/4$. Then the difference
Eq.\ \ref{*322} is constructed. 
Clearly after just a few blocking steps this operator
difference is close to zero. Similar results are observed in
other channels and give strong evidence of the existence of
a fixed point in the model. Combined with the previous
demonstrations regarding $\gamma_s$ and $d_H$ it appears that
the decimation method is a rather promising approach to 
a real-space RG for pure gravity.

\subsection{Higher derivative curvature term}

How robust is the ND method?  To answer that we have studied the
RG flow of other operators which are thought to be {\it irrelevant}
for pure gravity. One such operator is $\;O=\sum_i \log{q_i}\;$ which
corresponds to a set of higher derivative curvature interactions.
The non-perturbative flows of this operator have been examined
previously in the context of the LGB method \cite{simray}.
\nopagebreak
\begin{figure}
\centering
\epsfxsize=3.9in \epsfbox{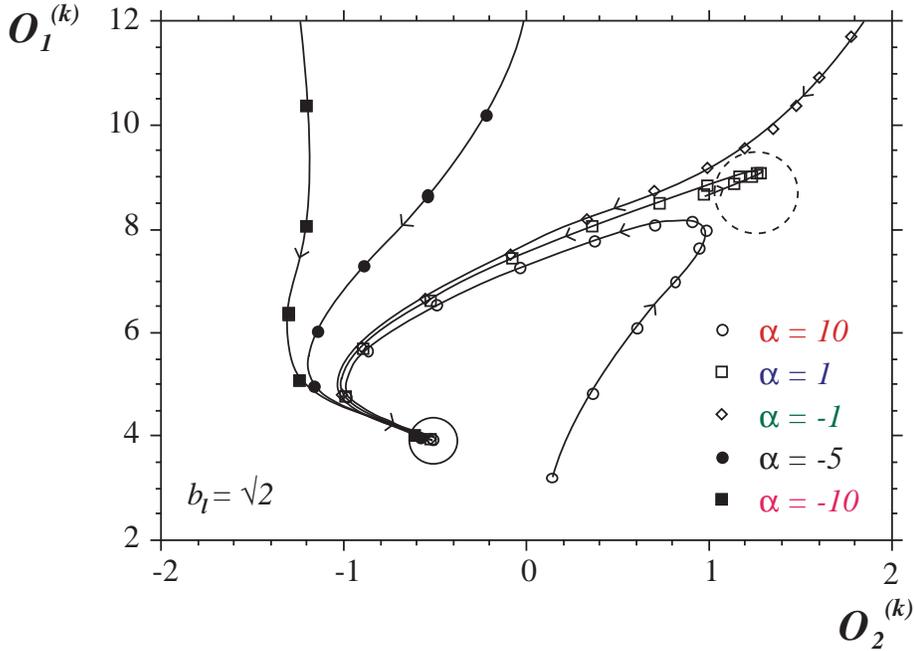}
\caption
 {The RG flow of the curvature operators $O_1$ and $O_2$
  as an ensemble of triangulations is blocked.  The
  flow is shown for different values of the bare
  coupling $\alpha$.  The dashed circle
  indicates an unstable fixed point, the solid a zero volume sink. }
\label{fig:5}
\end{figure}
\nopagebreak
The idea is to add such a term to the bare action with
some coupling $\alpha$. Under blocking the
coupling to this new term should be driven to zero in the effective
action as it corresponds to an irrelevant operator. Thus
expectation values of operators should approach their values
in the model without this extra interaction term.
This is illustrated in Fig.\ \ref{fig:5} were we plot the
RG trajectory in the ($O_1$,$O_2$) plane  
for a range of bare couplings $\alpha$.

We see that while the flows for different bare 
$\alpha$ start out from different points in the plane, for
sufficiently positive $\alpha$ they are all
attracted towards a fixed point. After passing close to
this point the flows converge on a unique `renormalized'
trajectory which ends at a trivial fixed point corresponding
to zero volume triangulations.
 
However, there are some indications
that this no longer is true for sufficiently large negative
values of $\alpha$ (the solid points). There the expectation values start to
deviate considerable from the pure gravity values. At these
lattice volumes they never pass close to the unstable fixed
point but run smoothly to the trivial end point. 
Such an effect has been observed before and
it is believed that for large negative $\alpha$ the model 
enters a `super-crumpled' state \cite{kaza}. 

\renewcommand{\tabcolsep}{7pt}

\begin{table}
\centering
\begin{tabular}{|cc|cccccc|} \hline
$k$ & $N^{(k)}$ & $\alpha=10$ & $\alpha=5$ &$\alpha=1$
& $\alpha=-1$ & $\alpha=-5$ & $\alpha=-10$  \\ \hline
0 &  1000  & -0.47(1)  & -0.47(2)  & -0.48(1) & -0.49(1)
  &  -0.03(3)  & 2.1(1)  \\
1 &  500   & -0.49(1)  & -0.47(2)  & -0.50(1) & -0.48(1)
  &  -0.30(2)  & 1.8(2)  \\
2 &  250   & -0.51(2)  & -0.46(2)  & -0.47(2) & -0.48(2)
  &  -0.53(2)  & 2.0(2)  \\
3 &  125   & -0.47(1)  & -0.42(2)  & -0.46(2) & -0.42(1)
  &  -0.65(3)  & 0.5(2)  \\ \hline
\end{tabular}
\caption{$\gamma_s^{(k)}$ for ensembles of triangulations obtained 
 by blocking $k$ times. Results are shown for various values
 of the bare coupling $\alpha$.}
\label{tab:2}
\end{table}

Such an effect is also observed in the fractal
structure.  We have measured the string susceptibility
exponent as before for several values of $\alpha$ and
various number of blocking steps.  The result is shown
in Table \ref{tab:2}.  For $10 \geq \alpha \geq -1$
we get consistently the pure gravity value.  But
for $\alpha = -5$ the initial ensemble
yields a different value of $\gamma_s$; but as it is
blocked the value appears to flow towards the
pure gravity fixed point value.  For $\alpha = -10$
the fractal structure is dramatically changed, 
we get a consistent value of $\gamma_s \approx 2$,
even after considerable blocking\footnote{The actual value of
$\gamma_s$ should not be taken too seriously as
the baby universe distribution for $\alpha=-10$ 
does not fit very
well the asymptotic form Eq.\ \rf{*42}. But
for a crumpled phase there is no reason to expect such a simple
asymptotic behavior, as has been
observed in simulations
of $4d$ simplicial gravity \cite{jan4d}.}.  

It is not clear at present whether these results can be explained
on the basis of finite size effects which are much more dramatic for
large negative $\alpha$ or whether they indicate that no
continuum limit exists for such models - the trajectories are
{\it never} attracted towards the usual unstable fixed point. 
Certainly we have not observed any non-trivial new fixed points 
associated with these negative coupling flows.

\section{Ising model coupled to 2{\it d} gravity}

The next step is to apply the method to matter fields
living on the surfaces.  For that purpose we chose
Ising spins coupled to gravity in which case the
matter contribution to the partition function 
Eq.\ \rf{*31} is 
\beq{*51}
Z_M(T) \;=\; \sum_{\{\sigma_i\}} {\rm e}^{
\textstyle \; \beta\sum_{\left<ij\right>}\sigma_i\sigma_j}.
\eeq
Here the sum runs over each edge length in a triangulation and the
coupling $\beta$ is tuned to its critical value.

This model is exactly solvable \cite{isol} so we know
the critical coupling, $\beta_c =  \frac{1}{2}\log{\frac{131}{85}}$, together
with the exponents governing the critical behavior.
In addition we know the effect the matter has on the
fractal structure.  
The Ising spins only change the 
string susceptibility exponent
at the critical point to
$\gamma_s = -1/3$, elsewhere it retains the pure
gravity value.

We have simulated this model at the critical point
of the Ising model for lattice volumes 
$N = 500, 1000$ and 2000 nodes.  
A Swendsen-Wang cluster algorithm
was used to update the spins and again about $10^6$
sweeps performed per lattice volume.

\subsection{Blocking the spin configurations}

When it comes to extending the ND method to
block a spin configuration, in addition to
a triangulation, we have to decide  
to what extent the blocking of the gravitational
sector and spin sector 
should influence each other. The simplest solution
is that the blocking of the triangulations should be 
performed precisely as for pure gravity - the only effect of
the critical matter will be to alter the initial distribution
of triangulations.  
When a 
given blocking ratio has been obtained a new
spin configuration is created from the spin
configuration of the initial triangulation\footnote{
We tried methods in which the surface and the spins
are blocked simultaneously.
After removing each node a new spin configuration
is found and in addition  
the possibility for choosing a given re-triangulation
is allowed to depend on
the spins.  These methods did not
seem to capture the physics of the system correctly.}.

To block the spin configuration we choose 
the "majority-rule" transformation that has been
applied successfully to the Ising model on a
regular lattice.  The spins on neighboring
nodes are grouped into "blocks" and a block spin value
assigned to each block depending on the sum of its spins.
If the sum is positive 
the value is +1 but -1 if it is negative. If the sum is zero
we assign +1 and -1 with equal probability.
However, there is a difference in applying this method to
random surfaces as opposed to a regular lattice. 
A regular lattice can be divided into equal
non-intersecting blocks, an arbitrary triangulation not.
Instead we define a "block" as a spin
that survives the decimation together with the neighbors
it had before blocking.  This definition is appealing as   
it incorporates the influences of the geometry.  The
size of the "spin-blocks" will be depend
on the local curvature. 

However, there is still an ambiguity in this definition of 
"spin-blocks".  Some spins will contribute to more
than one such block, in particular if two 
neighboring nodes both survive the decimation.
This is especially true if a small blocking ratio
is used.  Thus it might be necessary to limit the
influence of such nodes by excluding them from other
blocks than their own. We tried this modified definition
of spin-blocks, but it did not change the results
appreciably.

\subsection{The fractal structure}

\begin{figure}
\centering
\epsfxsize=4.7in \epsfbox{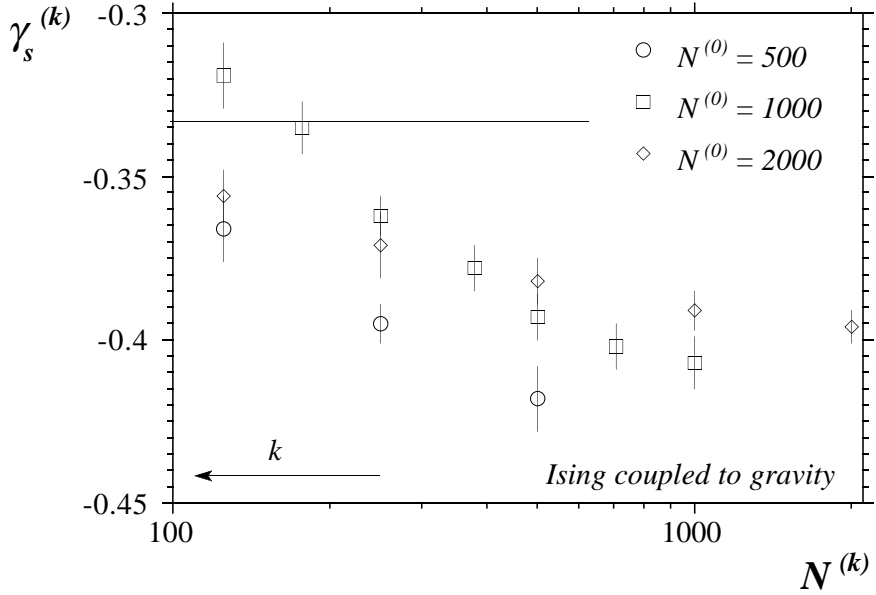}
\caption
 {Changes in $\gamma_s$ as triangulations with
  Ising spins are blocked. Data is shown for three
  initial volumes. Again the number of blocking steps $k$
  increases to the left.  The horizontal line is the
  exact value $\gamma_s = -1/3$. }
\label{fig:6}
\end{figure}

How well does the node decimation preserve the fractal
structure for this model?  Again we have
computed the string susceptibility exponent; the result
is shown in Fig.\ \ref{fig:6}
for three different initial volumes.
Notice that as the blocking of the geometry is independent
of the spins the results do not depend on the definition
of spin-blocks.

The first thing to notice is that without
any blocking we measure the value $\gamma_s \approx -0.41$, 
which is somewhat lower than the theoretical value
$\gamma_s = -1/3$.  This is due to the fact that we measure 
at the infinite volume critical coupling
$\beta_c^\infty$ not
the pseudo-critical coupling $\beta_c^V$ appropriate for 
the finite volume used.  This effect has been observed
in earlier numerical estimates of $\gamma_s$ for this
model in \cite{baby2}. There a sharp peak was observed
around $\beta_c^V$ with peak value $\gamma_s \approx -1/3$. 
But away from the peak the values rapidly return to the
pure gravity value $\gamma_s = -1/2$.  As the volume
is increased the peak moves closer to the true
critical point.

However under blocking the measured
$\gamma_s$ appears to flow towards the exact value.
This, of course, is very encouraging
from the point of view of capturing fixed point physics.
Notice that
direct simulations of the smallest volume lattices would have
produced estimates for $\gamma_s$ {\it further} from the exact result.

\subsection{Scaling exponents}

To measure the critical exponents of the Ising model
we consider several operators constructed out of
products of neighboring
spins and apply the methods described in section 2.
In practice the possible operators can be classified according 
to their symmetry properties under the $Z(2)$ operation $\sigma_i\to
-\sigma_i$.  In the linearized matrix $T_{\alpha \beta}$ the
cross-correlations between these even and odd operators
vanish.

In the odd sector we have looked at the operators
\beq{*55}
S^o_1 \;=\; \sum_i \sigma_i \;\;\;\;\;
{\rm and} \;\;\;\;\; 
S^o_2  \;=\; \sum_{\left<ijk\right>} \sigma_i \sigma_j \sigma_k,
\eeq
where the first sum is over nodes and the second over
triangles $\left<ijk\right>$.  For the even sector
we have chosen the operators
\beq{*56}
S^e_1 \;=\; \sum_{\left<ij\right>} \sigma_i \sigma_j \;,
\;\;\;\;\; S^e_2 \;=\; \sum_{\left<kl\right>^\prime}
\sigma_k \sigma_l \;\;\;\;\; {\rm and}
\;\;\;\;\; S^e_3 \;=\; \sum_{\left<ijkl\right>}
\sigma_i \sigma_j \sigma_k \sigma_l .
\eeq
The indices $i,j,k,l$ refer to an elementary link $ij$ together
with the complementary vertices $k,l$ associated with the two
neighboring triangles. 

These operators allow us to determine numerically two
scaling exponents; the magnetic exponent $y_1$ and 
the thermal exponent $y_2$. The former 
corresponds to the odd operators, the latter to
the even.  These scaling exponents are related to
the critical exponents of the Ising model
$\nu$ and $\delta$ through the relations
\beq{*57}
y_1  \;=\; \frac{\delta}{\delta + 1}
\;\;\;\;\;\; {\rm and}
\;\;\;\;\;\; y_2 \;=\; \frac{1}{\nu d_H}.
\eeq
For the Ising model coupled to gravity these critical 
exponents are known, $\nu d_H = 3$ and $\delta = 5$,
yielding the values $y_1 = 5/6$ and $y_2 = 1/3$ for
the scaling exponents.  In comparison, for the 
Ising model on a flat regular lattice these exponents
are $y_1 = 15/16$ and $y_2 = 1/2$.

\begin{figure}
\centering
\epsfxsize=4.7in \epsfbox{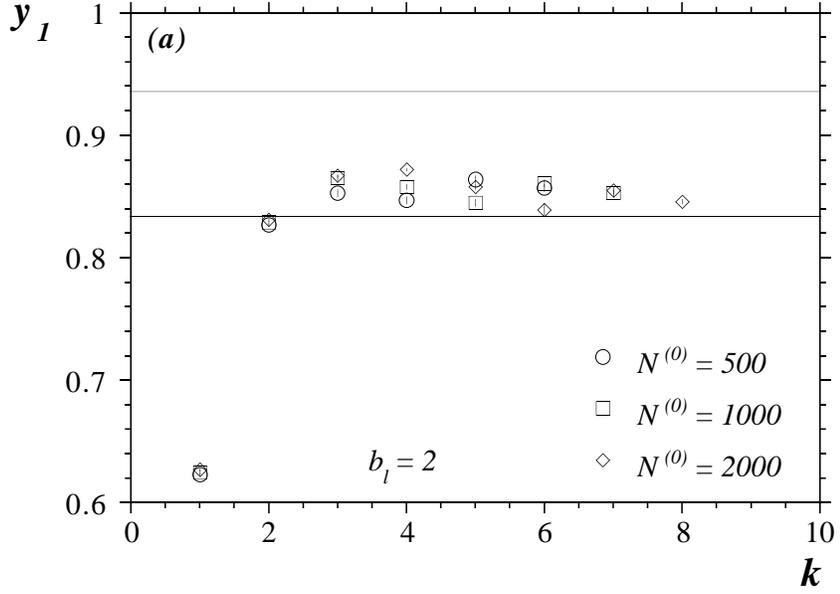}
\epsfxsize=4.7in \epsfbox{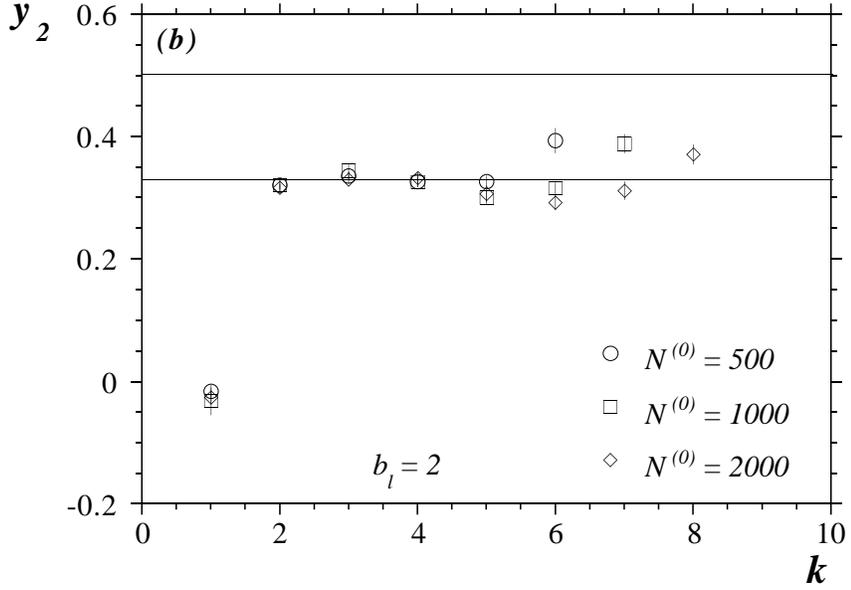}
\caption
 {The scaling exponents $y_1$ and $y_2$, corresponding to
  odd and even spin operators respectively, versus the number
  of RG iterations $k$. In both plots the lower
  horizontal lines are the exact values for
  an Ising model coupled to gravity $y_1 = 5/6$ and
  $y_2 = 1/3$, while the upper lines are  
  the flat space values $y_1 = 15/16$ and 
  $y_2 = 1/2$. }
\label{fig:7}
\end{figure}

In Figs.\ \ref{fig:7}$a$ and $b$ we show the extracted values of the
scaling exponents as a function of the RG iterations $k$ using a
blocking ratio $b_l = 2$ (and for
different initial volumes).  The values shown are obtained 
including only {\it one} spin operator for each sector in
the calculation of the matrix $T_{\alpha \beta}$, i.e.\
$S^o_1$ and $S^e_1$.  
With the accuracy we have at present no 
obvious improvement was observed by enlarging the operator basis
(although the results were perfectly compatible).

The results show that after one or two blocking steps we
have approached a critical fixed point. It clearly 
is the correct one as we get results which agree with 
the exact values 
within 2 or 3\% accuracy.  
In comparison the exponents for the Ising model on a
flat lattice are clearly ruled out. It is impressive
that even after blocking down to triangulations 
consisting only of 16 nodes the critical exponents
of the Ising model still come out correctly. 

Do these results depend on the blocking ratio?  Unlike for 
the triangulations the blocking of the spin configurations
is not independent of the amount of node decimation.  We tried three
different blocking ratios, $b_l = \sqrt{2}$, 2 and 4, but
the results did not change very much.
The error-bars increased somewhat as the blocking
ratio was decreased, which is not unexpected.

\section{Alternative blocking methods}

As mentioned in the introduction there are other real-space
RG transformations that have been proposed for 
random surfaces.  But how well do those methods
capture the long distance physics of the surfaces?
We have investigated this for the local geodesic
blocking (LGB) method.

A brief description of the method given in \cite{renken}: 
Given a
triangulation $T(N)$ we chose a subset of $N^\prime$ 
nodes and connect them together to form a block
triangulation $T^\prime$.  This is done in such way
as to preserve {\it locally}  
the relative geodesic distances between block nodes.  
In practice the subset of nodes is fixed at the
beginning of the simulations. The initial triangulation
is then evolved using a standard link flip algorithm and 
simultaneously the
block triangulation is updated 
with the prescription that a link $l_{ij}$ is flipped to
$l_{kl}$ if the geodesic distance $d_{kl}$ is less than
$d_{ij}$ (measured on the unblocked lattice).  That this 
constraint is local is essential for the algorithm to
be practical in numerical simulations.

We have measured the distribution of baby universes on the
ensemble of triangulations created by this blocking method,
both for surfaces with 
spherical and toroidal topology\footnote{Measuring $\gamma_s$ 
for manifolds of toroidal topology is slightly more
complicated than for a sphere.  Firstly, the minimal loop
used to identify a baby universe must not be topologically
non-trivial.  Secondly, the baby universe 
distributions now describe spheres growing on a torus.
Hence there are two {\it different} $\gamma_s$ exponents in Eq.\
\rf{*42}, one for each topology.  For pure
gravity both these exponents come out correctly in
simulations.}
(the latter was used in \cite{renken}).  The result for
$\gamma_s$ are shown in
Table 3.  They indicate that the fractal
structure is {\it not} correctly preserved.  The ensemble of
blocked triangulations has, consistently, $\gamma_s \approx 0.28$
independent of blocking ratio or topology.

\renewcommand{\tabcolsep}{8pt}

\begin{table}
\centering
\begin{tabular}{|cc|cc|cc|cc|}  \hline
&& \multicolumn{2}{c|}{$k = 0$} 
 & \multicolumn{2}{c|}{$k = 1$} 
 & \multicolumn{2}{c|}{$k = 2$}\\ 
&  $b_l$  &  $N$  &  $\gamma_s$  &  $N$  &  $\gamma_s$  
 &  $N$  &  $\gamma_s$  \\ \hline
{\it Sphere} &&&&&&& \\
&  1.08  &  270   &  -0.504(14)  &  250  &  0.271(18) && \\
&  1.2   &  300   &  -0.506(13)  &  250  &  0.280(11) && \\
&  1.6   &  400   &  -0.486(16)  &  250  &  0.296(27) && \\
&  2.0   &  500   &  -0.484(17)  &  250  &  0.273(30) && \\
&  4.0   &  1000  &  -0.499(11)  &  250  &  0.289(40) && \\ \hline
{\it Torus} &&&&&&& \\
&  4  &  2304   &  -0.486(16)  &  576  &  0.306(27) 
      &  144    &  0.281(30) \\
&  4  &  1024   &  -0.504(9)   &  256  &  0.268(35) && \\ \hline
\end{tabular}
\label{tab:3}
\caption{Measured values of $\gamma_s$ for triangulations
 blocked with the LGB method.  Results are shown for both 
 spherical and toroidal topologies and for various
 blocking ratios $b_l$.   }
\end{table}

This is in our opinion a serious flaw of the method.  If
$\gamma_s$ is thought of as a regular critical exponent this
result would imply that this RG scheme is driving the system 
towards a {\it new} fixed point - not pure two-dimensional gravity.
A possible explanation is that as the method
uses local constraints to dictate the link flips on the block
triangulations, it may actually simulate the effect of
having some kind of matter field living on the manifolds.
This could effectively change the physics of the model.

The situation might be improved by using a 
generalization of the LGB method; a
global geodesic blocking (GGB). It is defined by
requiring that the block triangulation preserves (up to
an overall global length rescaling) as closely as possible
{\it all} possible geodesic paths between block nodes.
In practice this may be implemented by minimizing a cost function $E$
when performing block lattice link flips
\beq{*62}
E=\sum_{ij}\left(d_{ij}-\lambda d^\prime_{ij}\right)^2.
\eeq
The quantities $d_{ij}$ and $d_{ij}^\prime$ are the distances
in units of the cut-off between block nodes $i$ and $j$ defined
on both the initial lattice and block lattice respectively. The
parameter $\lambda$
relates the two scales and may be chosen so as to minimize $E$ for
fixed lattices. Clearly other choices of $E$ are possible
corresponding, for example, to weighting small and long geodesics
differently. 

We have investigated this method numerically, using the
cost function Eq.\ \rf{*62}
and found that it indeed seems to cure the problem
of an incorrect $\gamma_s$ exponent. The values of 
$\gamma_s$, obtained with limited statistics, are
consistent with the pure gravity value. 
Unfortunately as the method is highly nonlocal it is
computationally burdensome and not feasible in practice.

\section{Outlook}

We have presented results on a new real-space renormalization group
method for dynamically triangulated $2d$ gravity. Our results 
are encouraging in so far as they indicate that
under a simple node decimation (ND)
flows toward a fixed
point in the critical surface are produced. Measurements of
the string susceptibility and Hausdorff dimension strongly
support the notion that this fixed point corresponds 
to $2d$ quantum gravity.  The effect of including an irrelevant
operator corresponding to a higher order curvature term 
also support this picture. Indeed the demonstrated {\it
irrelevance} of such an operator outside of perturbation
theory might be taken as the first successful non-trivial application
of the method. 

We have extended the method to a simple critical Ising system
coupled to $2d$ gravity and have been able to extract exponents 
in excellent agreement with the predictions
for gravitational dressing given by the DDK/KPZ formula \cite{DDK}. 
Furthermore measurements of the string susceptibility exponents
indicate that the finite size effects are significantly reduced.
This provides further evidence for a flow to the correct
fixed point.  

There is still some scope for improving the 
method.  A more systematic investigation of how to
inter-disperse the node and spin 
blocking would be desirable.  Also the effects of
employing a more extensive operator basis should be studied.

Provided the method or refinements thereof survive these
tests the way is open to apply these RG methods to a variety
of unsolved problems in $2d$ gravity.

\vspace{0.5in}
{\bf Acknowledgements}

We would like to acknowledge stimulating conversations during
the course of this work with Mark Bowick, Varghese John, John Kogut
and Ray Renken. This work was supported by research funds
provided by Syracuse University and the computational
facilities of NPAC.

\end{document}